\documentclass[preprint,12pt]{elsarticle}

\usepackage{amssymb}
\usepackage{amsmath}

\usepackage{array}
\usepackage{float}
\usepackage{graphicx}
\usepackage[margin=1in]{geometry} 
\usepackage{subcaption}

\usepackage{amsmath}

\usepackage{hyperref}

\usepackage{tikz}
\usetikzlibrary{shapes.geometric, arrows}

\usepackage{tabularx}
\usepackage{booktabs}

\usepackage{multirow}

\tikzstyle{startstop} = [rectangle, rounded corners, minimum width=3cm, minimum height=1cm,text centered, draw=black, fill=red!30]
\tikzstyle{process} = [rectangle, minimum width=3cm, minimum height=1cm, text centered, draw=black, fill=blue!30]
\tikzstyle{decision} = [diamond, minimum width=3cm, minimum height=1cm, text centered, draw=black, fill=green!30]
\tikzstyle{arrow} = [thick,->,>=stealth]

\journal{Computers in Biology and Medicine}

\begin{document}

\begin{frontmatter}



\title{\textbf{Impact of Shoe Parameters on Gait Using Wearables}}


\author[label1]{Nadeera Meghapathirana} 
\author[label1]{Oshada Rathnayake} 
\author[label2]{Thisali S Rathnayake}
\author[label1]{Roshan Godaliyadda} 
\author[label1]{Parakrama Ekanayake} 
\author[label1]{Vijitha Herath}
\affiliation[label1]{organization={Department of Electrical and Electronic Engineering},
            addressline={University of Peradeniya}, 
            city={Peradeniya},
            country={Sri Lanka}}

\affiliation[label2]{organization={Department of Electrical Engineering},
            addressline={University of Moratuwa}, 
            city={Moratuwa},
            country={Sri Lanka}}

\begin{abstract}
The study of biomechanics during locomotion provides valuable insights into the effects of varying conditions on specific movement patterns. This research focuses on examining the influence of different shoe parameters on walking biomechanics, aiming to understand their impact on gait patterns. To achieve this, various methodologies are explored to estimate human body biomechanics, including computer vision techniques and wearable devices equipped with advanced sensors. Given privacy considerations and the need for robust, accurate measurements, this study employs wearable devices with Inertial Measurement Unit (IMU) sensors. These devices offer a non-invasive, precise, and high-resolution approach to capturing biomechanical data during locomotion. Raw sensor data collected from wearable devices is processed using an Extended Kalman Filter to reduce noise and extract meaningful information. This includes calculating joint angles throughout the gait cycle, enabling a detailed analysis of movement dynamics. The analysis identifies correlations between shoe parameters and key gait characteristics, such as stability, mobility, step time, and propulsion forces. The findings provide deeper insights into how footwear design influences walking efficiency and biomechanics. This study paves the way for advancements in footwear technology and contributes to the development of personalized solutions for enhancing gait performance and mobility.
\end{abstract}




\end{frontmatter}



\section{Introduction}
\label{Introduction}
The biomechanics of human gait is a complex interplay of internal and external factors that contribute to stability, efficiency, and comfort during walking. The internal factors which affect the gait pattern can be listed as a person's health, sense of balance and coordination, age, and pathological asymmetry in the body \cite{Singhe}. Among the external factors such as surface type, incline/decline, weather condition, and physical environment, the kind of footwear plays a crucial role, in influencing the kinematics and kinetics of gait\cite{Cikajlo2008}. In particular, the height of a shoe's heel is known to significantly alter walking patterns, potentially leading to variations in posture, joint loading, and overall gait mechanics \cite{Cowley2009}, \cite{Stefanyshyn}, \cite{Zeng2023-qr}, \cite{PENG2021104355}.

Different shoe parameters, such as heel/platform height, ankle support, and shoe size have various effects on a person's gait pattern. Understanding these variations of gait is essential, not only for advancing the field of biomechanics but also for preventing women from using heels which affect them in the long term, and reducing the risk of musculoskeletal injuries \cite{Maduabuchi2012}. High heels alter natural biomechanics, leading to shortened strides, increased forefoot pressure, and changes in posture that shift the center of gravity forward \cite{Lee2001}. This unnatural positioning places excessive strain on the lower back, knees, and calf muscles, contributing to chronic pain and conditions such as plantar fasciitis and knee osteoarthritis \cite{sylvia2018biomechanical}. Long-term use can also cause shortening of the Achilles tendon and imbalance in muscle activation, reducing ankle mobility and increasing fall risk \cite{pannell2012postural}.

Several methods have been developed to analyse gait patterns, each with its own strengths and limitations. Traditional methods include vision-based systems, such as video motion capture and optical tracking, which rely on cameras to record and analyse movements \cite{VUN202495}. These methods are highly effective in controlled environments, offering detailed spatial data and the ability to visually assess gait \cite{Muro-de-la-Herran2014-fn}. However, they require sophisticated setups with multiple cameras located at various angles and heights, controlled lighting conditions, and can be subject to occlusion, making them less practical in real-world scenarios \cite{Hofmann}. It is also difficult to use in crowded places as well as when the subject is moving \cite{BOLDO2024108101}. There is also a concern in the privacy of the people who are monitored.

In contrast, wearable devices have emerged as a powerful alternative for gait analysis, offering a range of benefits over vision-based methods. Wearable technologies, such as inertial measurement units (IMUs), accelerometers, and gyroscopes, can be easily integrated into everyday clothing or attached directly to the body \cite{Tamura}. These devices provide continuous, real-time data, allowing for the analysis of gait patterns in naturalistic settings outside the laboratory \cite{XIANG2024108016}. This portability and ease of use make wearable devices particularly valuable for long-term monitoring and for analysing gait, in environments where traditional vision-based systems would be impractical, \cite{Benson}. Moreover, wearable devices can capture subtle variations in gait that may not be detected by vision-based methods such as movements of smaller joints, providing detailed insights into movement dynamics \cite{Tao2012}. Additionally, wearable technology enables researchers to study biomechanics within the body during locomotion\cite{WANG2006601}. As a result, wearable devices offer flexibility, accuracy, and practicality.  

IMUs are essential sensors used in applications ranging from navigation and stabilization in aerospace, automotive, and marine systems to motion tracking in robotics \cite{ZHI2022110516}, wearables \cite{XIANG2024108016} and healthcare \cite{GARCIAESTEBAN2024107935}. IMUs have become a popular choice for wearable devices in gait pattern analysis, offering the ability to capture detailed motion data in real-world environments \cite{Iosa2016} \cite{JIANG2022105905}. However, the accuracy of IMU-based measurements is often challenged by sensor noise, drift, and other sources of error that can compromise the reliability of gait analysis \cite{Park2021}. To address these challenges, Kalman filters are widely employed to estimate the true state of the system by optimally combining sensor data with a predictive model \cite{Ferdinando}. Traditional approaches, such as the Extended Kalman Filter (EKF), have been widely used to address this challenge. The EKF relies on linearizing the nonlinear functions of the system around the current estimate using a first-order Taylor expansion \cite{Gobbo2001}.

Among the various factors influencing gait, heel height, and platform height have been shown to significantly alter walking mechanics, affecting parameters such as step time and acceleration, and joint angles\cite{Barrera}. These parameters are crucial for understanding how different shoe designs impact gait and, consequently, the musculoskeletal health of the wearer. Temporal and spatial gait parameters, such as step time and acceleration, are foundational metrics that describe the rhythm of walking. Kinematic parameters, including joint angles at the knee, and ankle, offer a detailed view of the movement dynamics during gait. High-heeled shoes typically increase the plantarflexion of the ankle, which in turn affects the alignment and movement of the entire lower limb \cite{Kim2013}. This altered kinematic profile can lead to increased loading on specific joints, potentially contributing to discomfort or injury over time. 
 
Maduabuchi et al. discuss the effects of high heels on gait parameters such as step length, cadence, and velocity in their research but are limited to 3 heel heights \cite{Maduabuchi}. Taking it one step further Jiangyinzi Shang et al. use 2 sets of thick and thin heels with 3 heels heights.\cite{Shang2020-nb}. While previous studies have extensively explored the effects of heel height on gait, there remains a gap in the literature concerning the systematic analysis of how the different heel heights combined with different platform heights, ranging from flat shoes to high heels affect the walking gait patterns of women.   

This study aims to leverage the capabilities of IMUs to analyze the walking gait patterns of women wearing shoes with varying shoe parameters. By focusing on key gait parameters, such as joint angle mobility, step cycle time, and acceleration, this research seeks to deepen our understanding of the biomechanical implications of heel height, ultimately contributing to preventing musculoskeletal injuries.

In summary, the contributions of this study are as follows.
\begin{enumerate}
    \item  We collect walking gait pattern data from 3 different women wearing 7 different shoes, using 12 IMU-based wearable sensors.
    \item  The proposed method processes data---noise and drift mitigation and handling nonlinearity using an Extended Kallman Filter and extracts gait parameters including joint angle mobility, step cycle time, and acceleration.
    \item  This study draws relations between the variations of the extracted gait parameters with both the heel height as well as the platform heights providing additional information on how the respective shoe parameters affect the walking gait pattern in women.
\end{enumerate}

The structure of the manuscript is organized as follows: Section \ref{Materials and methods}: materials and methodology outline the device used, data collection method along with data filtering and analyzing. Section \ref{Results}: Results draw relations between gait parameters and shoe parameters. Section \ref{Discussion}: Discussion discusses the impact of shoe parameters on joint angle mobility, step cycle time, and acceleration. Finally, section \ref{Conclusions}: Conclusions summarize the findings of this study briefly and emphasize the importance of customizing footwear parameters to suit personal needs.

\section{Materials and methods}
\label{Materials and methods}
\subsection{\textbf{Wearable device technology for human gait analysis }}

Wearable devices have significantly advanced the monitoring of gait with high precision. Modern devices, often incorporating IMUs, pressure sensors, and flexible electronics, capture detailed data on human motion. Wearable devices are essential for human gait analysis because they provide continuous, real-time data, enabling monitoring in both laboratory and natural environments \cite{patel2012wearable}. Their portability allows for capturing natural human motions in everyday settings \cite{muro2014gait}. These devices also facilitate personalized assessments for rehabilitation and early diagnosis of gait abnormalities \cite{caldas2017human}. Additionally, Wearables offer a significant privacy advantage compared to traditional motion capture systems, such as computer vision systems\cite{del2016free}. While computer vision systems often rely on cameras and external sensors that capture and process large amounts of visual data, wearables focus solely on the user's movements through embedded sensors. This approach minimizes the risk of capturing sensitive or unintended visual information, offering a more secure and user-focused alternative for motion tracking. These Wearable devices are lightweight, non-invasive, and can be integrated into clothing or footwear, making them ideal for clinical, sports, and everyday settings. Machine learning algorithms further enhance data interpretation, providing insights into gait abnormalities and improving rehabilitation strategies.
IMUs which include accelerometers, gyroscopes, and magnetometers, are widely used in wearable devices. IMUs are particularly advantageous over other sensors and methods in human gait analysis due to their comprehensive motion capture capabilities, portability, and cost-effectiveness. Unlike optical motion capture systems, which require a controlled environment and are expensive, IMUs are compact, can be worn easily, and provide continuous, real-time data in natural settings. While pressure mats only measure foot pressure, IMUs capture detailed information on acceleration, angular velocity, and orientation, offering a more complete analysis of gait dynamics \cite{sabatini2006quaternion,tao2012gait}. Current research on IMU wearable device technology for human gait analysis focuses on the accuracy, miniaturization, and real-time processing of gait parameters.
Accordingly, this research focuses on analyzing human gait parameters, and an IMU-based wearable device was selected to capture motion data, given its advantages in accurately monitoring and analyzing human movements.

\subsection{\textbf{Wearable device}}

The wearable device is designed to capture real-time data in both indoor and outdoor environments. The design emphasizes wearability and flexibility. The modular architecture of the device allows for expansion and reduction with additional sensors and modules as needed for various applications. The initial modules exclusively utilize IMU sensing units due to their cost-effectiveness, portability, and minimally invasive solution for Human Motion Analysis (HMA), providing accurate and comprehensive data on human movement \cite{[07]useofIMUs}.

The device is designed with one central unit and two modules for each limb, both hands and both legs. Each hand module is equipped with four MPU 6050 IMU sensors, while each leg module contains six. The MPU 6050 is a 6-axis IMU sensor that integrates a 3-axis accelerometer and a 3-axis gyroscope, providing comprehensive motion data for precise tracking and analysis.

\begin{itemize}
    \item Central Unit: The device's central unit consists of an Arduino Nano board, a Real-Time Clock (RTC), and a power bank. It monitors and synchronizes data flow from the sensors, serves as the main power source, and stores data on an SD card. This ensures all connected modules operate in sync and maintain proper data recording and storage.
    
    \item Modules: Each module includes an Arduino Nano board, an SD card unit, and up to six sensors. These modules independently collect and store sensor data, reducing the computational load on the central unit. The design is modular, expandable, and wearable, with adjustable straps to secure sensors on various body parts, offering versatility for different applications and users.
    
	\item Circuit Design: The device operates on a 5V power supply, supporting the microcontrollers, SD card units, and IMU sensors. It features a start/stop button for data capture, an error button to mark important events, and an LED indicator on the Arduino Nano board for SD card status.
 
	\item Coding and Data Collection: Each module operates independently, typically collecting data in a loop format. Limiting each module to six sensors allows a higher data acquisition frequency than non-modular systems. The central unit synchronizes sensor readings across modules using trigger signals and an inbuilt timer, maintaining a sampling frequency of 32 Hz. LED indicators show the SD card status, including ready, start/stop, capturing, and error states. To conserve power, sensors enter sleep mode when not in use, and temperature sensors are disabled.
\end{itemize}
\subsection{\textbf{Sensor placement for measurements}}
\begin{figure}[H]
    \centering
    \begin{subfigure}[b]{0.25\linewidth}
        \centering
        \includegraphics[width=0.525 \linewidth]{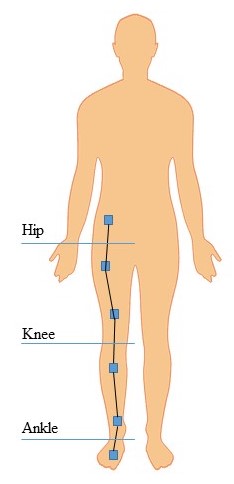}
        \caption{}
        \label{fig:Sensor placement}
    \end{subfigure}
    \begin{subfigure}[b]{0.25\linewidth}
        \centering
        \includegraphics[width=0.525 \linewidth]{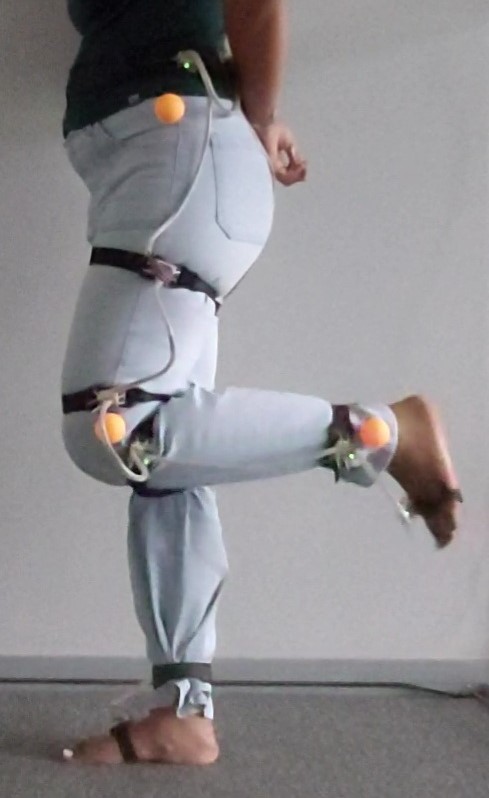}
        \caption{}
        \label{fig:sensorPlacementReal}
    \end{subfigure}
    \caption{Sensor placement}
    \label{fig:mainfigure1}
\end{figure}

The device can be used to capture whole-body movement sensor data. Since the focus was on comparing gait patterns and calculating gait parameters, the modules on both legs were used to obtain the measurements.
To achieve the goal of capturing data from three joint angles of legs: the ankle, knee, and hip joints. IMU sensor modules were set up as two sensor modules as on either side of a joint as shown in Fig.\ref{fig:Sensor placement}.

\subsection{\textbf{Data collection}}
\begin{figure}[H]
    \centering
    \begin{subfigure}[b]{0.5\textwidth}
        \centering
        \includegraphics[width=\textwidth]{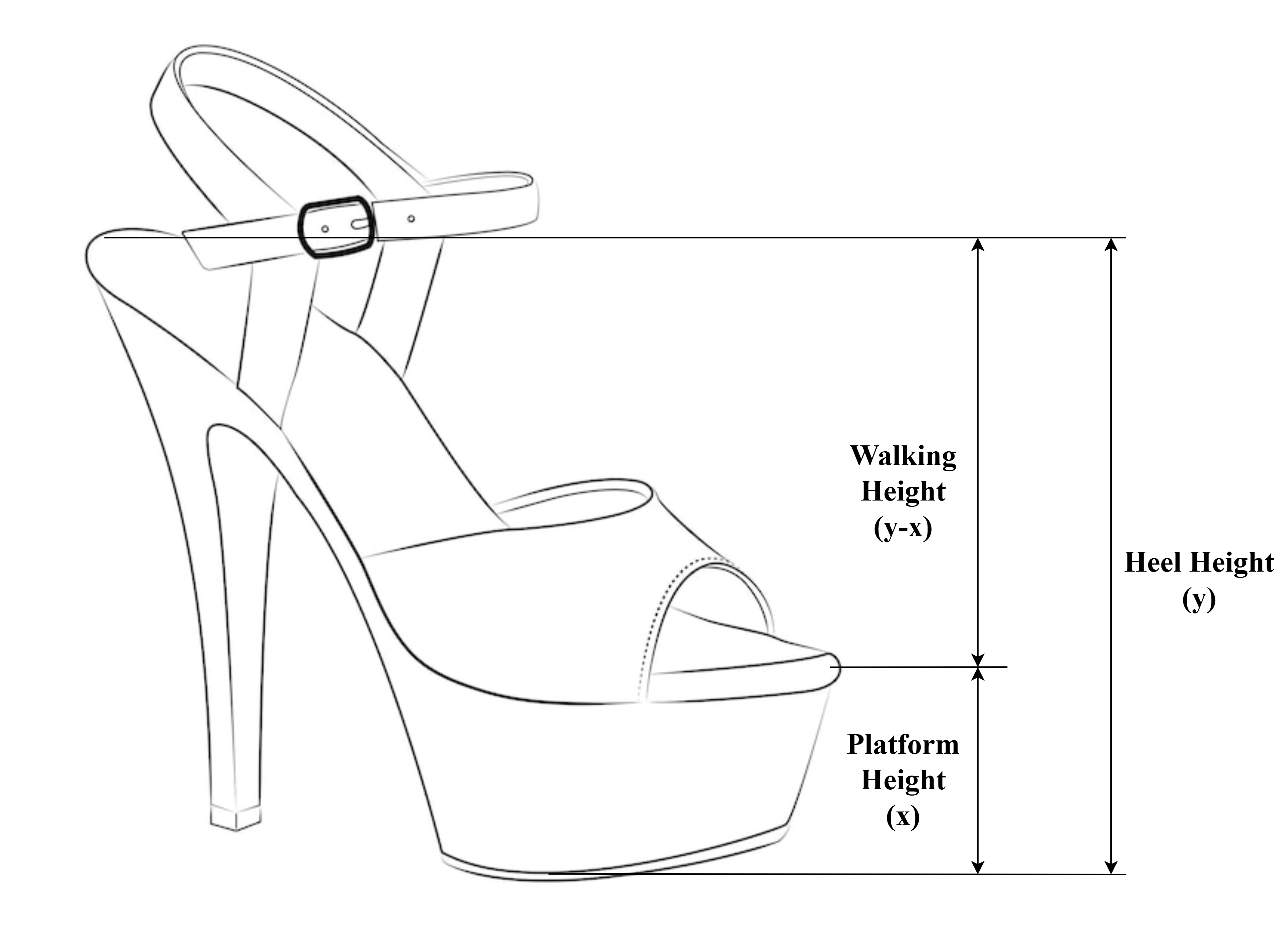}
        \caption{}
        \label{fig:Shoe Parameters}
    \end{subfigure}
    \hfill
    \begin{subfigure}[b]{0.575\textwidth}
        \centering
        \includegraphics[width=\textwidth]{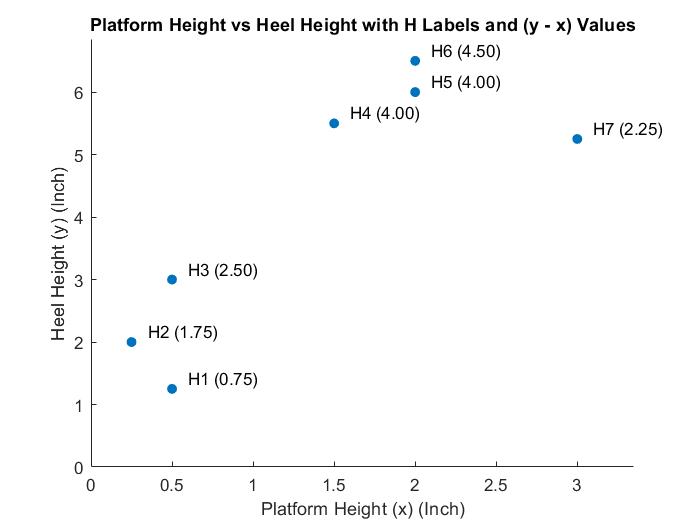}
        \caption{}
        \label{fig:Shoe Parameter Distribution}
    \end{subfigure}
    \caption{(a) Illustration of shoe parameters, including platform height(x), heel height(y), and walking height (y-x) and (b) Distribution of shoe parameters, showing the relationship between platform height (x-axis) and heel height (y-axis), with labels indicating specific shoe configurations (H1–H7)}
    \label{fig:mainfigure}
\end{figure}

\begin{itemize}
\item Candidates: The candidates were three healthy female adult volunteers aged 26 to 30 years, Height of about  165.1± 3.56 m and a Body mass:  45 - 65 kg.

\item Locomotion: Each participant was instructed to walk at a moderate speed of 20 steps per cycle while wearing shoes that have different height combinations of heels, platforms, and the device on flat terrain. The process was repeated 10 times to complete a total of 200 steps. This approach allowed for a comprehensive assessment of walking patterns across various conditions.

\item Shoe samples: Different shoe samples were used to examine the impact of different shoe parameters and shoe parameters are demonstrated in Fig.\ref{fig:Shoe Parameters}. All the shoes were lightweight and had the same physical design except the heel height, platform height, and walking height.
As in Fig.\ref{fig:Shoe Parameter Distribution} the shoe parameter distribution can be grouped into three clusters: H1, H2, and H3 in one cluster; H4, H5, and H6 in another; and H3 and H7 in the third. These clusters were formed to test the effects of walking height, platform height, and overall height, respectively.
\end{itemize}

\subsection{\textbf{Data analysing}}
Analyzing stepping acceleration, step time, and joint angles calculated from raw IMU data provides a comprehensive view of gait mechanics. Stepping acceleration reflects the dynamic forces exerted during the gait cycle, capturing propulsion, deceleration, and ground reaction impacts, which are critical for identifying irregularities or asymmetries in movement. Step time measures the temporal rhythm of each step, offering insights into stability, and fatigue, while joint angles provide detailed kinematic data on joint flexion, extension, and rotations, highlighting biomechanical efficiency and range of motion. Together, these parameters enable precise evaluation of gait patterns and the influence of factors such as footwear on stability, comfort, and performance.

Kalman filtering is a powerful tool used in denoising raw data and sensor fusion, particularly for transforming raw data from IMUs into useful information like rotational angles. By combining the strengths of different sensors and applying the Kalman filter, raw IMU data can be transformed into accurate and stable estimates of rotational angles, enabling precise tracking of orientation. The Kalman filter effectively reduces noise from sensor measurements, compensates for the drift in gyroscope measurements using accelerometer and magnetometer data, and provides real-time estimation of rotational angles. Considering the nonlinear characteristics of human movement, this study utilizes the Extended Kalman Filter (EKF) \cite{grewal2001kalman,kalman1960new}. 

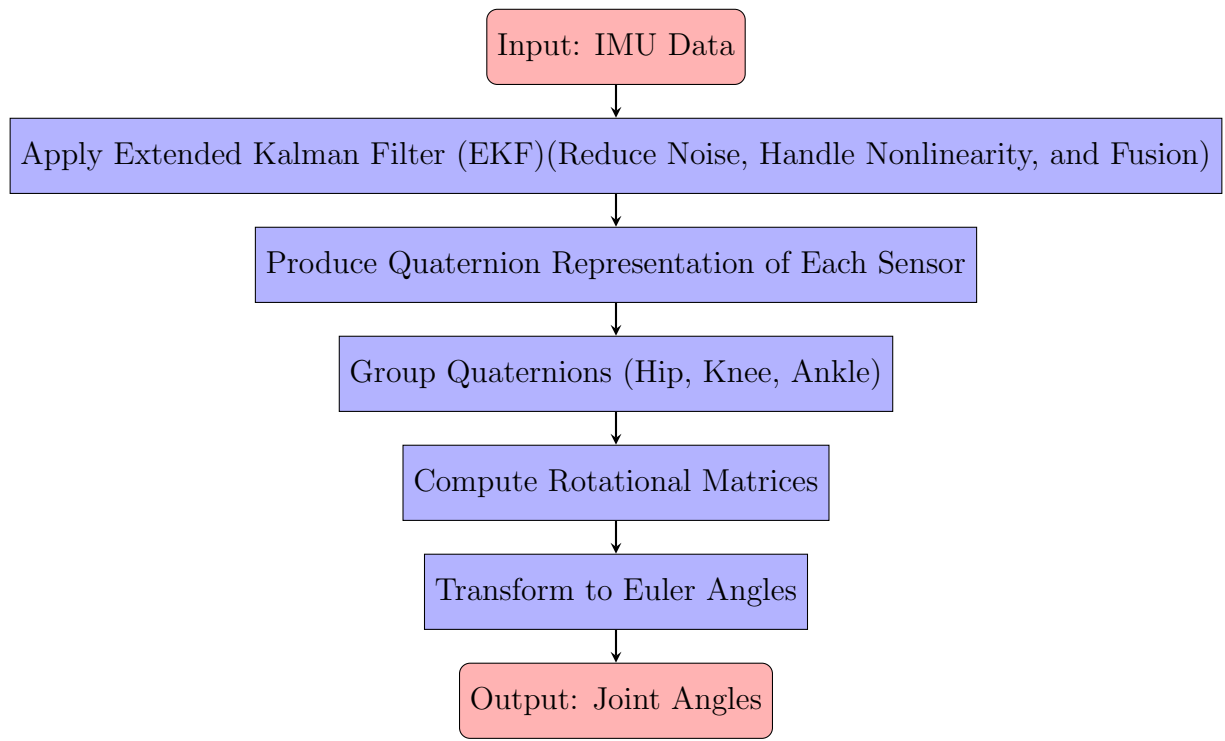
\begin{figure}[H]
\centering
\begin{tikzpicture}[node distance=1.45cm]

\node (start) [startstop] {Input: IMU Data};
\node (ekf) [process, below of=start] {Apply Extended Kalman Filter (EKF) \newline (Reduce Noise, Handle Nonlinearity, and Fusion)};
\node (quaternions) [process, below of=ekf] {Produce Quaternion Representation of Each Sensor};
\node (grouping) [process, below of=quaternions] {Group Quaternions (Hip, Knee, Ankle)};
\node (matrices) [process, below of=grouping] {Compute Rotational Matrices};
\node (euler) [process, below of=matrices] {Transform to Euler Angles };
\node (end) [startstop, below of=euler] {Output: Joint Angles};
\draw [arrow] (start) -- (ekf);
\draw [arrow] (ekf) -- (quaternions);
\draw [arrow] (quaternions) -- (grouping);
\draw [arrow] (grouping) -- (matrices);
\draw [arrow] (matrices) -- (euler);
\draw [arrow] (euler) -- (end);    
    
\end{tikzpicture}
\caption{Flowchart Illustrating the Extended Kalman Filtering Process for Sensor Fusion and Joint Angle Estimation.}
\label{fig:kalman_flowchart}
\end{figure}

Extended Kalman Filter is an extension of the standard Kalman Filter that can be used to handle nonlinear systems effectively and predict human gait patterns. The EKF can work with multiple sensors such as gyroscopes and accelerometers, providing a more accurate estimation of the gait parameters by reducing noise. As demonstrated in Fig.\ref{fig:kalman_flowchart}, the denoising process produces quaternion representations for each sensor relative to a global reference frame \cite{li2002survey}. Since there are six sensors, respective six quaternions are then grouped into three pairs—one for the hip, one for the knee, and one for the ankle. Each pair is used to compute a rotational matrix that describes the rotation between the two sensors. Ultimately, these rotation matrices are transformed into Euler angles, which represent the angles at each joint.

\subsubsection{Joint angle variation}
\label{sec:joint angles}
\begin{figure}[H]
    \centering
    \includegraphics[width=0.8\linewidth]{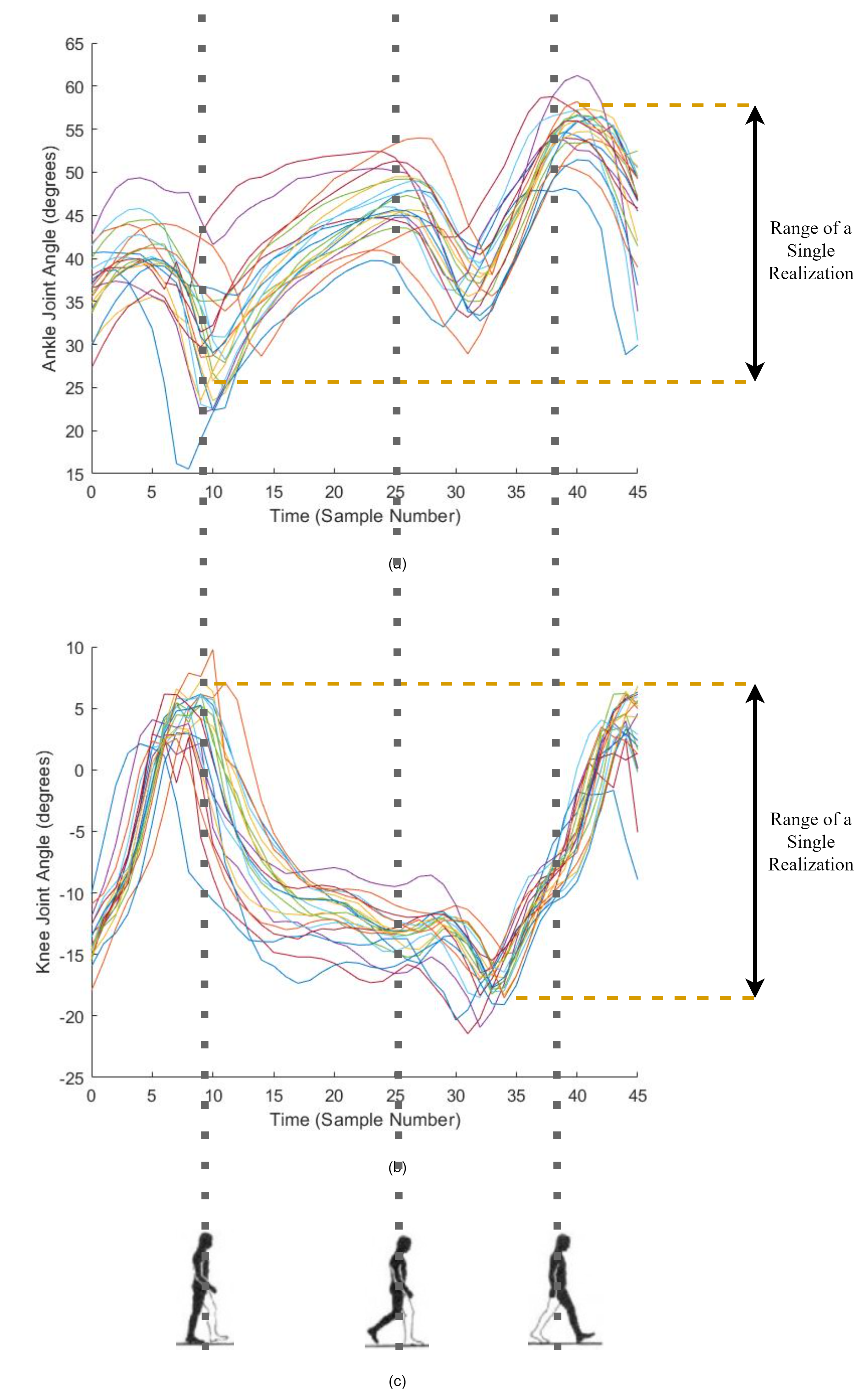}
    \caption{Joint Angle Variation During a Gait Cycle (Referring to the Black Colored Leg) (a)Ankle Angle Variation. (b) Knee Angle Variation. (c) The Gait Cycle}
    \label{fig:Joint Angle Variation During a Gait Cycle}
\end{figure}

Fig.\ref{fig:Joint Angle Variation During a Gait Cycle} illustrates the ankle and knee joint angle variations over a gait cycle for a single candidate wearing one type of shoe. Each plot represents multiple realizations, showcasing how the joint angles fluctuate across individual steps.

To thoroughly analyze the variation in joint angles, the maximum and minimum values for each joint across all recorded realizations were examined. The range of a specific realization from a ankle, and knee angle variation plot is presented in Fig.\ref{fig:Joint Angle Variation During a Gait Cycle}. This approach allows to obtain a set of range values for each individual joint angle. Once these range values are collected, a detailed statistical analysis is conducted for each joint angle to gain deeper insights into their variability and characteristics. Analysis was conducted based on the shoe parameter distribution illustrated in Fig.\ref{fig:Shoe Parameter Distribution}
\subsubsection{Step cycle time}
\label{sec:step cylcle}
Step Cycle Time, also referred to as the gait cycle or stride time, is the time taken for a complete gait cycle, from the point when one foot contacts the ground until the same foot contacts the ground again. This parameter is essential for understanding the rhythm and regularity of an individual’s walking pattern, as it directly relates to walking speed, balance, and stability.

Since walking is a repetitive movement, the sensor data also has a repetitive pattern, and the step time is given by measuring the time between key points---peak points of IMU data from a selected sensor. A sensor that is near the ankle is considered because it has a higher range of motion. In addition, the mean step time is calculated to avoid the effect of inconsistent walking patterns.  

\subsubsection{Acceleration}
\label{sec:acc}
The acceleration magnitude is a critical gait parameter that represents the overall intensity of motion by combining the three-dimensional accelerations (x, y, z axes). It is mathematically calculated as the Euclidean norm: 
The acceleration magnitude is calculated using the following formula:

\[
\text{Acceleration  magnitude} = \sqrt{x^2 + y^2 + z^2}
\]
where the x, y, and z represent the acceleration along the respective axis, with respective to the sensor frame. Before this calculation, the raw acceleration data from the accelerometer is filtered using the EKF.
 Moreover, the mean acceleration magnitude is a significant parameter in gait analysis as it provides a scalar representation of the combined dynamic forces during walking or running, making it useful for understanding the overall physical exertion or activity level. Additionally, variations in acceleration magnitude over time can highlight gait abnormalities, inconsistencies, or changes due to external factors such as footwear or terrain. Moreover, steady and periodic patterns in acceleration magnitude are indicative of a stable gait, whereas irregularities may suggest instability or balance issues \cite{moe2004trunk, kavanagh2008accelerometry, zijlstra2003spatio}. 


\section{Results}
\label{Results}

\begin{figure}[H]
    \centering
    \begin{subfigure}{0.3275\textwidth}
        \includegraphics[width=\textwidth]{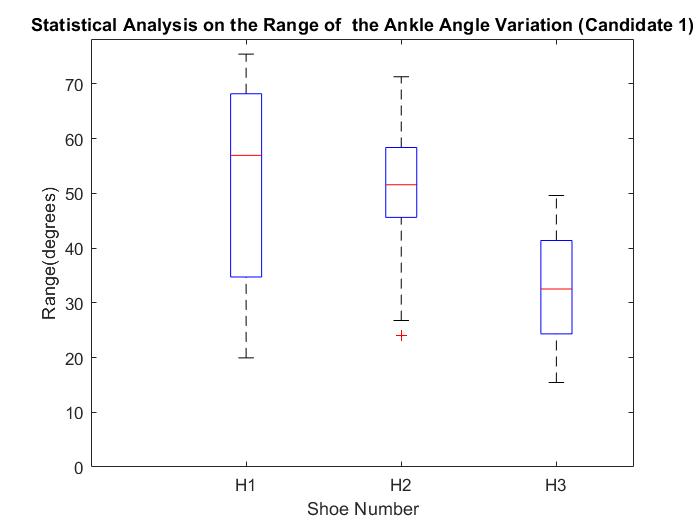}
        \caption{} 
        \label{fig:HeelandAnkle_1}
    \end{subfigure}
    \hfill
    \begin{subfigure}{0.3275\textwidth}
        \includegraphics[width=\textwidth]{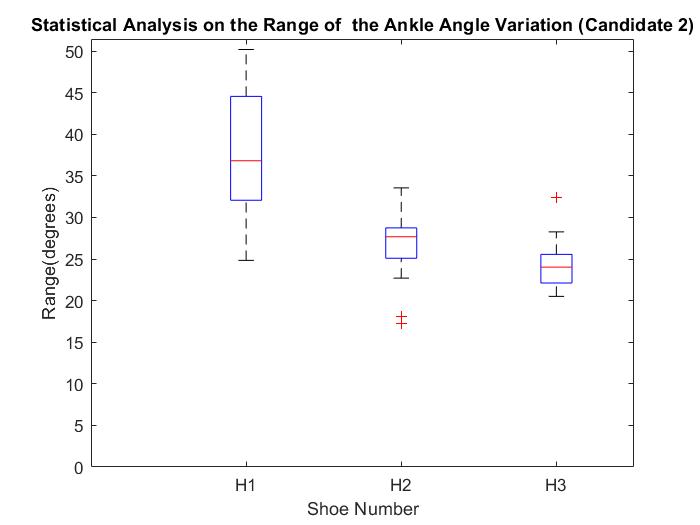}
        \caption{} 
        \label{fig:HeelandAnkle_2}
    \end{subfigure}
    \hfill
    \begin{subfigure}{0.3275\textwidth}
        \includegraphics[width=\textwidth]{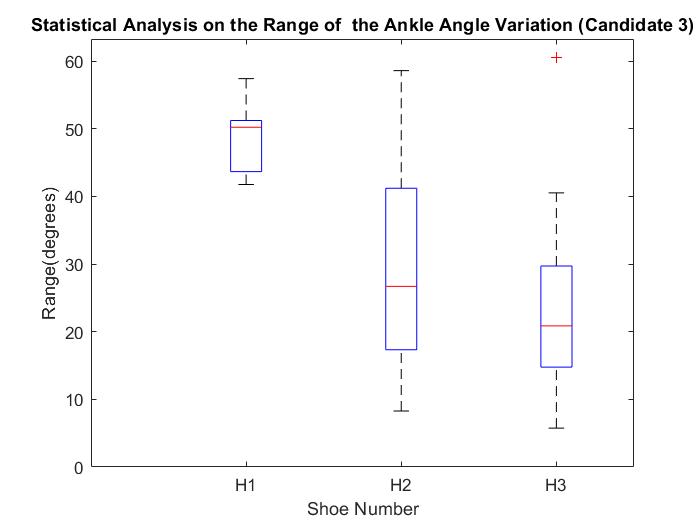}
        \caption{} 
        \label{fig:HeelandAnkle_3}
    \end{subfigure}
    
    \vspace{0.5cm} 

    \begin{subfigure}{0.3275\textwidth}
        \includegraphics[width=\textwidth]{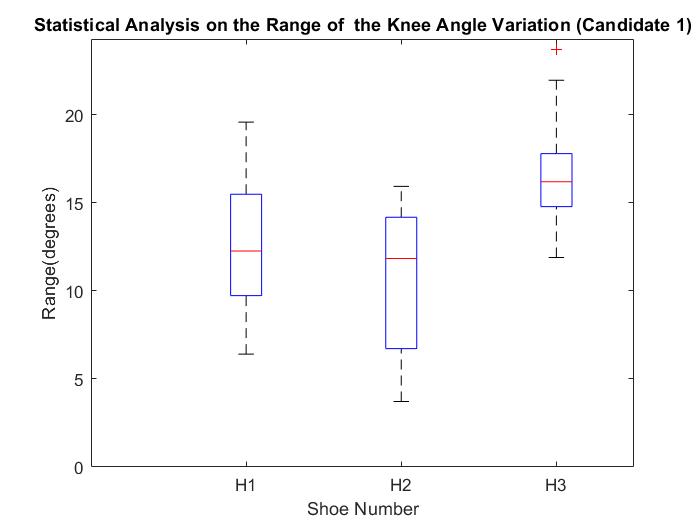}
        \caption{} 
        \label{fig:walking_Heght_knee_1}
    \end{subfigure}
    \hfill
    \begin{subfigure}{0.3275\textwidth}
        \includegraphics[width=\textwidth]{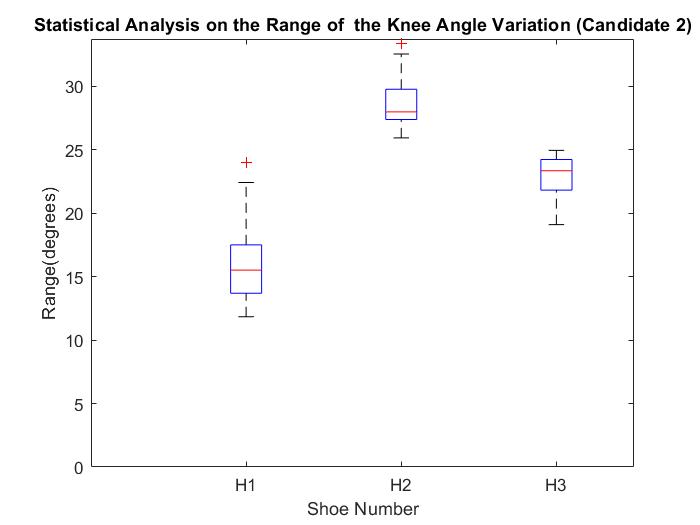}
        \caption{} 
        \label{fig:walking_Heght_knee_2}
    \end{subfigure}
    \hfill
    \begin{subfigure}{0.3275\textwidth}
        \includegraphics[width=\textwidth]{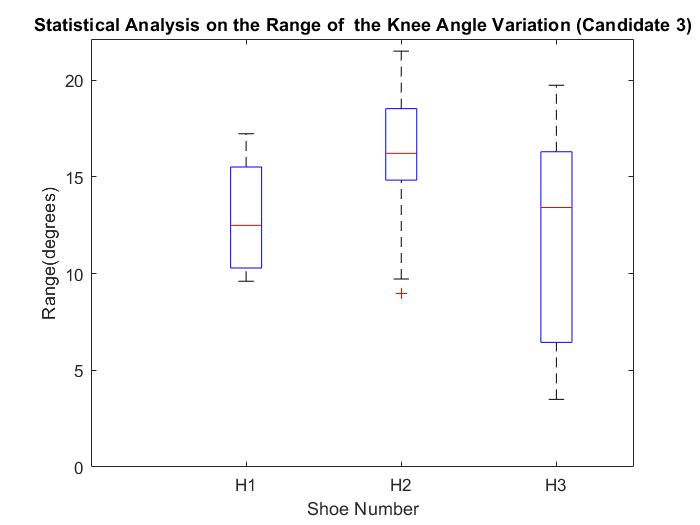}
        \caption{} 
        \label{fig:walking_Heght_knee_3}
    \end{subfigure}
    
    \vspace{0.5cm} 

    \begin{subfigure}{0.3275\textwidth}
        \includegraphics[width=\textwidth]{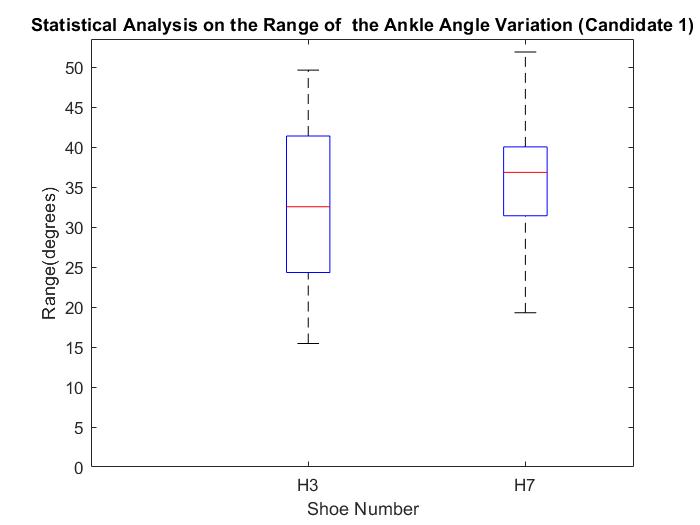}
        \caption{} 
        \label{fig:Platform_Effect_ankle_1}
    \end{subfigure}
    \hfill
    \begin{subfigure}{0.3275\textwidth}
        \includegraphics[width=\textwidth]{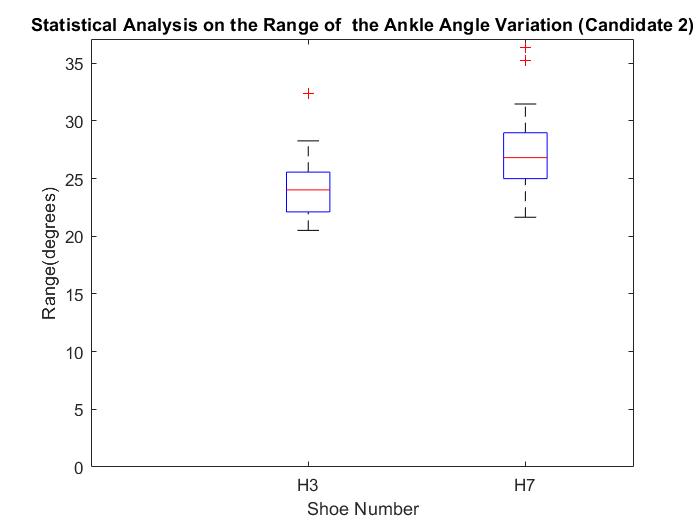}
        \caption{} 
        \label{fig:Platform_Effect_ankle_2}
    \end{subfigure}
    \hfill
    \begin{subfigure}{0.3275\textwidth}
        \includegraphics[width=\textwidth]{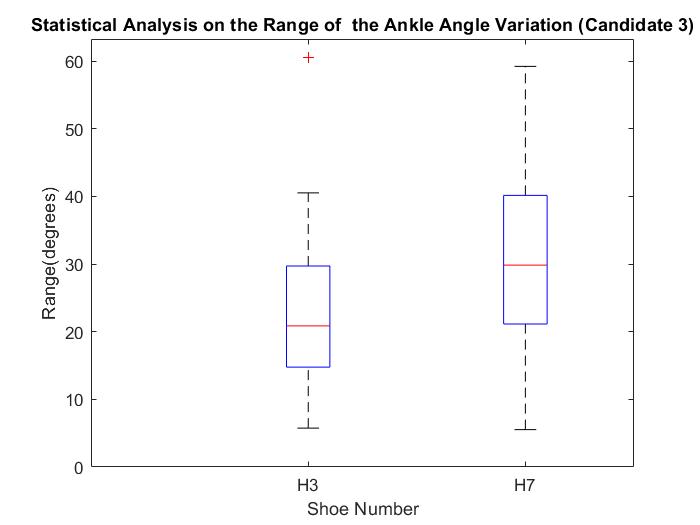}
        \caption{} 
        \label{fig:Platform_Effect_ankle_3}
    \end{subfigure}
    
    \vspace{0.5cm} 

    \begin{subfigure}{0.3275\textwidth}
        \includegraphics[width=\textwidth]{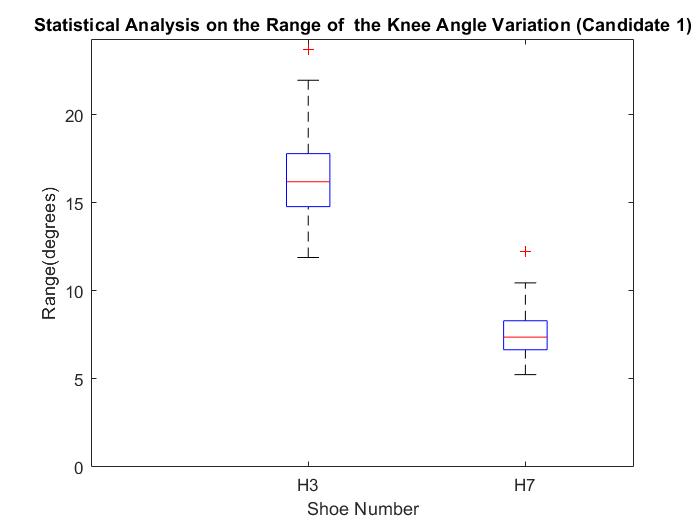}
        \caption{} 
        \label{fig:Platform_Effect_knee_1}
    \end{subfigure}
    \hfill
    \begin{subfigure}{0.3275\textwidth}
        \includegraphics[width=\textwidth]{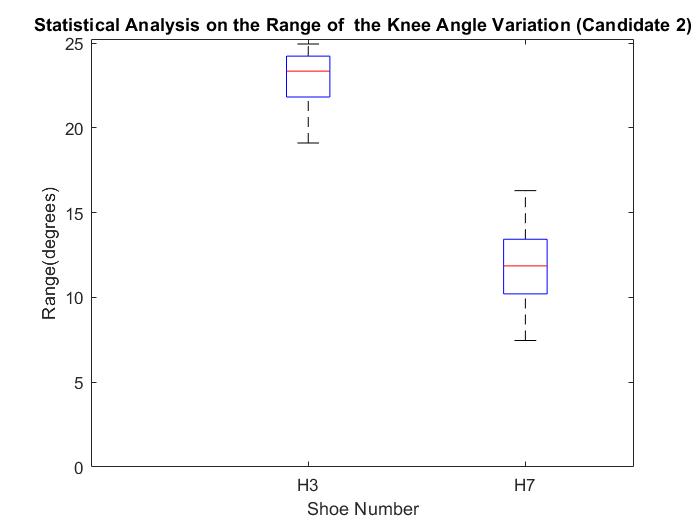}
        \caption{} 
        \label{fig:Platform_Effect_knee_2}
    \end{subfigure}
    \hfill
    \begin{subfigure}{0.3275\textwidth}
        \includegraphics[width=\textwidth]{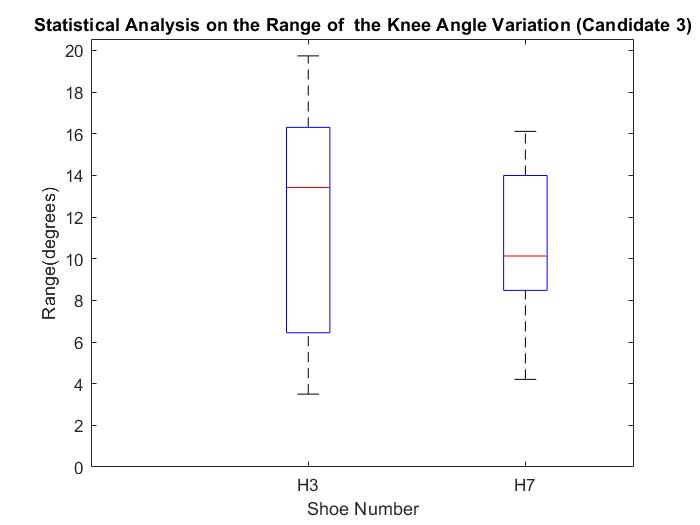}
        \caption{} 
        \label{fig:Platform_Effect_knee_3}
    \end{subfigure}

    \caption{Statistical analysis of the range of the joint angle variation. (a)-(c) represents the ankle angle variation of the three candidates for three sets of shoe parameters (H1, H2, and H3). (d)-(f) highlights the knee angle variation of the three candidates for three sets of shoe parameters (H1, H2, and H3). (g)-(i) demonstrate the ankle angle variation of the three candidates for two sets of shoe parameters (H3, and H7), and (j)-(l) showcase the knee angle variation of the two candidates for three sets of shoe parameters (H3, and H7).}
    \label{fig:12figures_labeled}
\end{figure}

The range of angle variations for ankle and knee joint was calculated as described in Section \ref{sec:joint angles}, for different shoe parameters. This analysis was further visualized through box plots, as shown in Fig. \ref{fig:12figures_labeled}.

The first row of box plots (Fig. \ref{fig:HeelandAnkle_1} - \ref{fig:HeelandAnkle_3})presents the statistical analysis of the ankle angle variation across three candidates, considering three different shoe parameters. The results indicate that the mean range of ankle angle variation decreases as the walking height (y-x) increases and it is common for all three candidates.

In contrast, the second row of plots (Fig. \ref{fig:walking_Heght_knee_1} - \ref{fig:walking_Heght_knee_3})focuses on the knee angle variation for the same walking heights (y-x). Unlike the ankle, there is no clear correlation between the mean range of knee angle variation and walking height (y-x).

The remaining plots analyze the impact of platform height on ankle and knee angle variations. For the knee joint, the mean range of angle variation decreases as platform height increases. Meanwhile, the ankle angle variation remains consistent with the trend observed for walking height (y-x).

Table 01 explores the relationship between shoe configurations, step cycle time, and mean acceleration magnitudes for three different candidates during walking trials. Key variables include shoe platform and heel heights, step cycle time, and the corresponding mean acceleration magnitudes across different shoe designs.Acceleration and step cycle time were calculated as mentioned in \ref{sec:step cylcle} and \ref{sec:acc} sections.

\begin{table}[h!]
\centering
\caption{Step cycle time, Mean acceleration magnitude, and Average variance for different Shoes with different parameters}
\resizebox{\textwidth}{!}{%
\begin{tabular}{|l|c|c|c|c|c|c|c|c|c|c|c|}
\hline
\multirow{2}{*}{\textbf{Shoe}} & \multicolumn{2}{c|}{\textbf{Shoe Parameters}} & \multicolumn{3}{c|}{\textbf{Step Cycle Time (s)}} & \multicolumn{3}{c|}{\textbf{Mean Acceleration(m/s$^2$)}} & \multicolumn{3}{c|}{\textbf{Variance of Acceleration(m/s$^2$)$^2$}} \\ 
\cline{2-12}
 & \textbf{Platform} & \textbf{Heel} & \multirow{2}{*}{\textbf{Cand.01}} & \multirow{2}{*}{\textbf{Cand.02}} & \multirow{2}{*}{\textbf{Cand.03}} & \multirow{2}{*}{\textbf{Cand.01}} & \multirow{2}{*}{\textbf{Cand.02}} & \multirow{2}{*}{\textbf{Cand.03}} &\multirow{2}{*}{\textbf{Cand.01}} & \multirow{2}{*}{\textbf{Cand.02}} & \multirow{2}{*}{\textbf{Cand.03}} \\

 & \textbf{(Inches)} & \textbf{(Inches)} & \textbf{} & \textbf{} & \textbf{} & \textbf{} & \textbf{} & \textbf{} & \textbf{} & \textbf{} & \textbf{} \\
 
\hline
H1 & 0.5 & 0.75 & 1.5555 & 1.1474 & 1.7432 & 1.6051 & 1.9650 & 1.6672 & 0.1236 & 0.3921 & 0.0712 \\ 
\hline
H2 & 0.25 & 2.0 & 1.6146 & 1.1183 & 1.8323 & 1.7169 & 2.0345 & 1.6322 & 0.1230 & 0.4477 & 0.1351 \\ 
\hline
H3 & 0.5 & 3.0 & 1.5512 & 1.1152 & 1.8673 & 1.8477 & 2.1798 & 1.7959 & 0.1266 & 0.5166 & 0.1047 \\ 
\hline
H4 & 1.5 & 5.5 & 1.5736 & 1.1875 & 1.5784 & 1.6476 & 2.1169 & 1.3863 & 0.1362 & 0.5074 & 0.1280 \\ 
\hline
H5 & 2.0 & 6.0 & 1.5552 & 1.1439 & 1.2492 & 1.5411 & 1.8944 & 1.6281 & 0.1283 & 0.5170 & 0.1439 \\ 
\hline
H6 & 2.0 & 6.5 & 1.5688 & 1.1710 & 1.2788 & 1.6262 & 2.1910 & 1.6254 & 0.1700 & 0.6039 & 0.1357 \\ 
\hline
H7 & 3.0 & 5.25 & 1.6152 & 1.1508 & 1.2492 & 1.3847 & 1.8944 & 1.5125 & 0.1288 & 0.5170 & 0.0992 \\ 
\hline
\end{tabular}
}
\label{tab:shoe_data}
\end{table}

The platform heights range from 0.25 inches (H2) to 3 inches (H7), and heel heights range from 0.75 inches (H1) to 6.5 inches (H6). Shoes with higher heels and platforms (e.g., H8, H9) tend to result in slower step cycles and reduced acceleration, indicating a trade-off between height and walking performance. Lower step cycle times are observed with shoes featuring smaller heel heights, such as H2 and H5, likely indicating greater walking efficiency. Conversely, higher heel heights, such as H6 and H7, result in increased step cycle times, suggesting more effort or reduced walking efficiency. Higher acceleration magnitudes are observed for mid-range heel heights (e.g., H3), indicating more dynamic motion or improved stability. Lower acceleration magnitudes are associated with higher heels, such as H7. Different candidates exhibit distinct walking patterns under the same shoe configurations, highlighting individual differences in gait and balance. Candidate 02 generally shows higher acceleration magnitudes compared to Candidates 01 and 03.

\section{Discussion}
\label{Discussion}
\subsection{Joint angle mobility}
Analyzing joint angles derived from raw acceleration and angular velocity data offers several benefits in observing and understanding gait patterns. By capturing movement at a granular level, this analysis allows for precise insights into the biomechanical behaviors and deviations associated with walking, running, and other forms of locomotion. Moreover, joint angles can reveal how gait patterns adapt to external influences, such as changes in terrain or footwear. This understanding is critical for designing interventions or products tailored to individual needs.

\subsubsection{Effect of the walking height} 
H1, H2, and H3 were analyzed to evaluate the impact of walking height (y-x) on ankle and knee angle variations throughout the gait cycle. The findings, illustrated in Fig.\ref{fig:HeelandAnkle_1},\ref{fig:HeelandAnkle_2}, and \ref{fig:HeelandAnkle_3}, show a clear trend: as walking height (y-x) increases, the range of motion in the ankle becomes progressively restricted across all participants. This indicates that elevated walking heights limit the natural flexibility and dynamic movement of the ankle joint, potentially altering gait mechanics.

In contrast, knee angle variation demonstrates a different pattern, as shown in Fig.\ref{fig:walking_Heght_knee_1},\ref{fig:walking_Heght_knee_2}, and \ref{fig:walking_Heght_knee_3}. Unlike the ankle, the knee does not exhibit a notable restriction in motion associated with increased walking height. However, a general decrease in the interquartile range is observed for both ankle and knee angles as the walking height (y-x) increases. This trend suggests that higher walking heights lead to a more consistent and stable walking pattern, likely due to reduced variability in joint movements under these conditions.

\subsubsection{Effect of the overall height (keeping the walking height nearly constant)}

Based on the analysis of H3 and H7, which share similar walking height (y-x) values but differ in platform height (x), the findings reveal key patterns regarding joint mobility. As shown in Fig. \ref{fig:Platform_Effect_knee_1},\ref{fig:Platform_Effect_knee_2}, and \ref{fig:Platform_Effect_knee_3}, the knee angle variation decreases as the overall elevation rises. This is evident in the box plots, where the range of knee motion becomes narrower with increased elevation, indicating restricted knee mobility. This suggests that participants adopt a more cautious gait as their overall height increases, likely to maintain balance and stability during walking. 

In contrast, the ankle angle variation, illustrated in Fig. \ref{fig:Platform_Effect_ankle_1},\ref{fig:Platform_Effect_ankle_2}, and \ref{fig:Platform_Effect_ankle_3}, continues to demonstrate a consistent trend with walking height (y-x) but remains unaffected by the overall elevation. The box plots show that the ankle's range of motion is predominantly influenced by the walking height (y-x), with minimal variation caused by differences in platform height. This distinction highlights the ankle's role in adapting to walking height changes while remaining less sensitive to the overall elevation.

\subsection{Step cycle time and acceleration}
The analysis of step time and acceleration provides critical insights into how variations in heel and platform heights influence gait dynamics and stability. Step cycle time reflects the efficiency and rhythm of walking, while acceleration magnitude and its variance reveal the effort and consistency in maintaining balance and forward motion. This exploration highlights the unique responses of each candidate, offering a understanding for evaluating walking efficiency and gait stability under varying conditions and adaptability of individual gait patterns.\\

The step cycle time data were analyzed under the following conditions: \\
\ref{(a)} Effect of Heel Height (Keeping Platform Height nearly constant) \\
\ref{(b)} Effect of Platform Height Variation (Keeping Heel Height nearly constant)\\
\ref{(C)} Effect of Both Heel and Platform Heights (Keeping walking height constant)\\

\subsubsection{Effect of heel height (keeping platform height constant)}
\label{(a)}
As the heel height increases from H1 (0.75') to H3 (3'), all candidates show a reduction in step cycle time. The step cycle time of the candidate 01, remains fairly consistent (1.5552s to 1.5512s) despite variations in heel height. This reflects a high level of adaptability to heel height changes. Candidate 02 shows a significant reduction in step cycle time, from 1.1474 s for H1 to 1.1152 s for H3, indicating, that the candidate achieves more efficient gait with increased heel height on a constant platform height. Candidate 03, similar to Candidate 01, exhibits minimal changes in step cycle time, ranging from 1.7432 s for H1 to 1.8673 s for H3, highlighting consistent walking patterns with slight adjustments to accommodate the varying heel height. This overall trend suggests that increased heel height, within the studied range, can enhance gait efficiency for certain individuals while maintaining stability.

Candidate 01 has shown a more moderate increase in acceleration (from 1.6051 m/s² to 1.8477 m/s²), which reflects better adaptability to changes in heel height. This suggests that Candidate 01 can efficiently adjust gait mechanics, maintaining a stable and consistent posture even with increased heel height. The lowest variance for Candidate 01 further supports the observation of smoother and more consistent gait patterns.

In contrast, the steady increase in average acceleration magnitude from 1.9650 m/s² (H1) to 2.1798 m/s² (H3) for Candidate 02 indicates that higher heel heights demand greater effort to maintain balance and forward motion. This rise suggests that Candidate 02 experiences more pronounced challenges with higher heels, which may slightly affect stability. The higher variance for Candidate 02 implies that their gait becomes less consistent with increased heel height. While the acceleration magnitudes increase, this variability may indicate difficulties in maintaining stable walking patterns or seamlessly adapting to heel height changes.

For Candidate 03, the relatively consistent acceleration magnitudes (1.6672 m/s² to 1.7959 m/s²) suggest that their gait is less affected by heel height changes. This steadiness highlights a more stable dynamic, possibly due to consistent stride mechanics or natural walking adjustments. The moderate variance for Candidate 03 indicates some variability, but overall, the patterns remain steady. While Candidate 03 adapts to heel height changes with reasonable stability, the adjustments are not as seamless as those observed in Candidate 01.

Sometimes, even when the average acceleration magnitude decreases with increasing heel height, the step cycle time for the same candidate may increase. This could indicate either a lower level of stability or a more cautious approach to adapting gait in response to the changes in heel height. A lower acceleration magnitude coupled with a longer step cycle time suggests that the candidate may be prioritizing balance and control over speed, potentially as a way to mitigate instability caused by the altered biomechanics of walking with higher heels. This cautious approach reflects an effort to maintain posture and stability, even if it results in a slower, less efficient gait. Conversely, this behavior may also highlight challenges in achieving efficient gait adjustments under such conditions, as the candidate requires more time to adapt to the dynamic changes introduced by increased heel height.

\subsubsection{Effect of platform height variation (keeping heel height nearly constant)}
\label{(b)}
H4 to H7 shoes have a negligible heel height difference, approximately around 1 inch. These shoes are suitable for analyzing the impact of platform height on gait acceleration and step cycle time.

With the increment of the platform height, the step cycle time of the candidate 01 has increased moderately from 1.5736s (H4) to 1.6152s (H7). The moderate increase in step cycle time shows that Candidate 02 experiences some effect of platform height changes but adapts well overall. The consistency in step cycle time reflects better postural control and stability under these conditions. As platform height increases, the step cycle time of the candidate 02, increases slightly, which could suggest a need for more effort to stabilize gait. Higher platforms may cause slight instability, requiring more time to complete each step. This could indicate less adaptability to changing terrain height compared to other candidates. Like Candidate 01, Candidate 03 shows only minor changes in step cycle time, indicating strong adaptability and stability. This trend highlights that platform height has a relatively low impact on the walking dynamics of this candidate.

The effect of platform height, with heel height kept nearly constant, reveals unique patterns in the average acceleration magnitude and its variance for each candidate. Candidate 02 exhibits a significant increase in average acceleration magnitude with platform height, peaking at 2.1910 m/s² for H6 and slightly reducing to 1.8944 m/s² at H7. This trend indicates that higher platform heights introduce greater challenges for Candidate 02, reflecting difficulty in maintaining balance and adapting to elevated surfaces. The high variance in acceleration magnitudes for Candidate 02 further emphasizes the inconsistency in their gait when platform heights increase. Candidate 01 demonstrates a more stable range of acceleration magnitudes, from 1.6476 m/s² (H4) to 1.3847 m/s² (H7). This stability suggests superior adaptability and control over gait mechanics, even as platform heights vary. The low variance in acceleration magnitudes for Candidate 01 supports this observation, indicating smoother and more consistent walking patterns across the conditions. Candidate 03 maintains relatively consistent acceleration magnitudes, ranging from 1.3863 m/s² to 1.6281 m/s², showing minimal impact from platform height variations. The moderate variance observed for Candidate 03 reflects a steady gait with minor adjustments, highlighting stable walking dynamics but slightly less adaptability compared to Candidate 01.

\subsubsection{Effect of both heel and platform heights (keeping walking height constant)}
\label{(C)}

The comparison of H3 and H7 shoes highlights the combined effect of platform and heel height on step cycle time as follows:
The minimal difference in step cycle time between H3 and H7 indicates that Candidate 01 is less affected by the combination of platform and heel heights. This reflects a stable walking pattern and an ability to adapt to combined changes in footwear parameters. The slightly slower step cycle time for H7 (3' platform and 5.25' heel height) compared to H3 (3' heel only) suggests that platforms introduce additional instability compared to heels alone. This may be due to the elevated base, which increases the risk of imbalance during walking. The nearly identical step cycle times between H3 and H7 show that Candidate 03 maintains consistent gait mechanics, even with variations in heel and platform heights. This demonstrates strong adaptability and controlled walking dynamics.

The effect of walking height (difference between heel height and platform height) is evident in the average acceleration magnitude and its variance for each candidate. For Candidate 02, the acceleration is higher for H3 (2.1798 m/s²) compared to H7 (1.8944 m/s²), suggesting that a constant heel height with a lower platform provides better stability and reduces the effort required for gait adjustments. However, the higher variance in acceleration for Candidate 02 reflects inconsistent gait patterns under these combined conditions, indicating challenges in adapting seamlessly to variations in walking height. Candidate 01 demonstrates consistent acceleration magnitudes, with a slightly higher value for H3 (1.8477 m/s²) compared to H7 (1.3847 m/s²). This suggests stable gait dynamics and efficient adaptation to changes in walking height. The lowest variance observed for Candidate 01 further confirms adaptability and consistent walking patterns under varying conditions. For Candidate 03, the acceleration magnitudes remain relatively steady, with minimal differences between H3 (1.7959 m/s²) and H7 (1.6281 m/s²). This reflects consistent gait mechanics and a steady adaptation to both heel and platform height variations. The moderate variance observed for Candidate 03 highlights a balanced gait with minor adjustments, indicating stability without significant disruptions.
\section{Conclusions}
\label{Conclusions}
The analysis of gait dynamics under varying heel and platform height conditions highlights key insights into the unique adaptation patterns and stability mechanisms demonstrated by each candidate. The findings emphasize the impact of footwear design, walking surfaces, and individual biomechanical responses on walking efficiency and stability.
Increasing walking height enhances step cycle efficiency due to biomechanical adjustments, such as improved forward propulsion mechanics and reduced ankle joint mobility. In addition, elevated overall height introduces instability, generally reducing gait efficiency and consistency, compelling individuals to walk more cautiously. Furthermore, the results show that while all candidates adapt to higher heels, the degree of effort and stability varies and personalized interventions are crucial. These can include customizing footwear (e.g., heel height and platform design) or providing targeted rehabilitation exercises to address specific gait issues. By aligning interventions with an individual's unique gait dynamics, it is possible to enhance balance and stability more effectively. This is particularly important for individuals with movement disorders, aging populations, or those recovering from injury. Gait instability or inefficient movement patterns can increase the risk of falls and other injuries. The findings suggest that personalized adjustments can mitigate these risks by improving the individual’s ability to adapt to changes in their footwear. Moreover, this study presents several promising avenues for future research, including the evaluation of the effects of sports equipment on optimal performance and the systematic monitoring of employee performance throughout the workday.

\bibliographystyle{elsarticle-num}  
\bibliography{ref}

\begin{thebibliography}{10}
\expandafter\ifx\csname url\endcsname\relax
  \def\url#1{\texttt{#1}}\fi
\expandafter\ifx\csname urlprefix\endcsname\relax\def\urlprefix{URL }\fi
\expandafter\ifx\csname href\endcsname\relax
  \def\href#1#2{#2} \def\path#1{#1}\fi

\bibitem{Singhe}
S.~Singh, R.~Bohara, A.~Rahman, R.~Indoria, Effect of different types of footwear on gait pattern, a scoping review 1 2 design: Scoping review 12 scholar databases with following study selection criteria: 14, International Journal of Design and Nature Volume 2 (02 2024).

\bibitem{Cikajlo2008}
I.~Cikajlo, Z.~Matjačić, The influence of boot stiffness on gait kinematics and kinetics during stance phase, Ergonomics 50 (2008) 2171--82.
\newblock \href {https://doi.org/10.1080/00140130701582104} {\path{doi:10.1080/00140130701582104}}.

\bibitem{Cowley2009}
E.~Cowley, T.~Chevalier, N.~Chockalingam, The effect of heel height on gait and posture a review of the literature, Journal of the American Podiatric Medical Association 99 (2009) 512--8.

\bibitem{Stefanyshyn}
D.~Stefanyshyn, B.~Nigg, V.~Fisher, B.~O'Flynn, W.~Liu, The influence of high heeled shoes on kinematics, kinetics, and muscle emg of normal female gait, Journal of Applied Biomechanics 16 (2000) 309--319.
\newblock \href {https://doi.org/10.1123/jab.16.3.309} {\path{doi:10.1123/jab.16.3.309}}.

\bibitem{Zeng2023-qr}
Z.~Zeng, Y.~Liu, X.~Hu, P.~Li, L.~Wang, Effects of high-heeled shoes on lower extremity biomechanics and balance in females: a systematic review and meta-analysis, BMC Public Health 23~(1) (2023) 726.

\bibitem{PENG2021104355}
Y.~Peng, D.~W.-C. Wong, T.~L.-W. Chen, Y.~Wang, G.~Zhang, F.~Yan, M.~Zhang, Influence of arch support heights on the internal foot mechanics of flatfoot during walking: A muscle-driven finite element analysis, Computers in Biology and Medicine 132 (2021) 104355.
\newblock \href {https://doi.org/https://doi.org/10.1016/j.compbiomed.2021.104355} {\path{doi:https://doi.org/10.1016/j.compbiomed.2021.104355}}.

\bibitem{Maduabuchi2012}
J.~Maduabuchi, N.~Joseph, A.~Egwuonwu, A.~O. Ezeukwu, C.~Nwafulume, Effects of different heel heights on selected gait parameters of young undergraduate females, Page Header Logo Journal of Paramedical Sciences 3 (2012) 2008--4978.

\bibitem{Lee2001}
C.-M. Lee, E.-H. Jeong, A.~Freivalds, Biomechanical effects of wearing high-heeled shoes, International Journal of Industrial Ergonomics 28 (2001) 321--326.
\newblock \href {https://doi.org/10.1016/S0169-8141(01)00038-5} {\path{doi:10.1016/S0169-8141(01)00038-5}}.

\bibitem{sylvia2018biomechanical}
A.~Sylvia, A biomechanical examination of the lower extremities in high heeled shoes, a (2018).

\bibitem{pannell2012postural}
S.~L. Pannell, U.~MD, The postural and biomechanical effects of high heel shoes: A literature review, Journal of Vascular Surgery (2012).

\bibitem{VUN202495}
D.~S.~Y. Vun, R.~Bowers, A.~McGarry, Vision-based motion capture for the gait analysis of neurodegenerative diseases: A review, Gait \& Posture 112 (2024) 95--107.
\newblock \href {https://doi.org/https://doi.org/10.1016/j.gaitpost.2024.04.029} {\path{doi:https://doi.org/10.1016/j.gaitpost.2024.04.029}}.

\bibitem{Muro-de-la-Herran2014-fn}
A.~Muro-de-la Herran, B.~Garcia-Zapirain, A.~Mendez-Zorrilla, Gait analysis methods: An overview of wearable and non-wearable systems, highlighting clinical applications, Sensors (Basel) 14~(2) (2014) 3362--3394.

\bibitem{Hofmann}
M.~Hofmann, S.~Sural, G.~Rigoll, Gait recognition in the presence of occlusion: A new dataset and baseline algorithms, in: 19th International Conference in Central Europe on Computer Graphics, Visualization and Computer Vision, WSCG 2011 - In Co-operation with EUROGRAPHICS, Full Papers Proceedings, 19th International Conference in Central Europe on Computer Graphics, Visualization and Computer Vision, WSCG 2011 - In Co-operation with EUROGRAPHICS, Full Papers Proceedings, 2011, pp. 99--104, 19th International Conference in Central Europe on Computer Graphics, Visualization and Computer Vision, WSCG 2011 ; Conference date: 31-01-2011 Through 03-02-2011.

\bibitem{BOLDO2024108101}
M.~Boldo, R.~{Di Marco}, E.~Martini, M.~Nardon, M.~Bertucco, N.~Bombieri, On the reliability of single-camera markerless systems for overground gait monitoring, Computers in Biology and Medicine 171 (2024) 108101.
\newblock \href {https://doi.org/https://doi.org/10.1016/j.compbiomed.2024.108101} {\path{doi:https://doi.org/10.1016/j.compbiomed.2024.108101}}.

\bibitem{Tamura}
T.~Tamura, Wearable Inertial Sensors and Their Applications, Academic Press, 2014, Ch. 2.2, pp. 85--104.
\newblock \href {https://doi.org/10.1016/B978-0-12-418662-0.00024-6} {\path{doi:10.1016/B978-0-12-418662-0.00024-6}}.

\bibitem{XIANG2024108016}
L.~Xiang, Y.~Gu, Z.~Gao, P.~Yu, V.~Shim, A.~Wang, J.~Fernandez, Integrating an lstm framework for predicting ankle joint biomechanics during gait using inertial sensors, Computers in Biology and Medicine 170 (2024) 108016.
\newblock \href {https://doi.org/https://doi.org/10.1016/j.compbiomed.2024.108016} {\path{doi:https://doi.org/10.1016/j.compbiomed.2024.108016}}.

\bibitem{Benson}
L.~Benson, C.~Clermont, E.~Bošnjak, R.~Ferber, The use of wearable devices for walking and running gait analysis outside of the lab: A systematic review, Gait \& Posture 63 (05 2018).
\newblock \href {https://doi.org/10.1016/j.gaitpost.2018.04.047} {\path{doi:10.1016/j.gaitpost.2018.04.047}}.

\bibitem{Tao2012}
W.~Tao, T.~Liu, R.~Zheng, H.~Feng, Gait analysis using wearable sensors, Sensors (Basel, Switzerland) 12 (2012) 2255--83.
\newblock \href {https://doi.org/10.3390/s120202255} {\path{doi:10.3390/s120202255}}.

\bibitem{WANG2006601}
W.~Wang, A.~D. Stefano, R.~Allen, A simulation model of the surface emg signal for analysis of muscle activity during the gait cycle, Computers in Biology and Medicine 36~(6) (2006) 601--618.
\newblock \href {https://doi.org/https://doi.org/10.1016/j.compbiomed.2005.04.002} {\path{doi:https://doi.org/10.1016/j.compbiomed.2005.04.002}}.

\bibitem{ZHI2022110516}
Z.~Zhi, D.~Liu, L.~Liu, A performance compensation method for gps/ins integrated navigation system based on cnn–lstm during gps outages, Measurement 188 (2022) 110516.
\newblock \href {https://doi.org/https://doi.org/10.1016/j.measurement.2021.110516} {\path{doi:https://doi.org/10.1016/j.measurement.2021.110516}}.

\bibitem{GARCIAESTEBAN2024107935}
J.~García-Esteban, B.~Curto, V.~Moreno, F.~Hernández, P.~Alonso, F.~Serrano, F.~Blanco, Real needle for minimal invasive procedures training using motion sensors and optical flow, Computers in Biology and Medicine 170 (2024) 107935.
\newblock \href {https://doi.org/https://doi.org/10.1016/j.compbiomed.2024.107935} {\path{doi:https://doi.org/10.1016/j.compbiomed.2024.107935}}.

\bibitem{Iosa2016}
M.~Iosa, P.~Picerno, S.~Paolucci, G.~Morone, Wearable inertial sensors for human movement analysis, Expert Review of Medical Devices 13 (06 2016).
\newblock \href {https://doi.org/10.1080/17434440.2016.1198694} {\path{doi:10.1080/17434440.2016.1198694}}.

\bibitem{JIANG2022105905}
Y.~Jiang, P.~Malliaras, B.~Chen, D.~Kulić, Real-time forecasting of exercise-induced fatigue from wearable sensors, Computers in Biology and Medicine 148 (2022) 105905.
\newblock \href {https://doi.org/https://doi.org/10.1016/j.compbiomed.2022.105905} {\path{doi:https://doi.org/10.1016/j.compbiomed.2022.105905}}.

\bibitem{Park2021}
S.~Park, S.~Yoon, Validity evaluation of an inertial measurement unit (imu) in gait analysis using statistical parametric mapping (spm), Sensors 21~(11) (2021).
\newblock \href {https://doi.org/10.3390/s21113667} {\path{doi:10.3390/s21113667}}.

\bibitem{Ferdinando}
H.~Ferdinando, H.~Khoswanto, D.~Purwanto, Embedded kalman filter for inertial measurement unit (imu) on the atmega8535, in: 2012 International Symposium on Innovations in Intelligent Systems and Applications, 2012, pp. 1--5.
\newblock \href {https://doi.org/10.1109/INISTA.2012.6246978} {\path{doi:10.1109/INISTA.2012.6246978}}.

\bibitem{Gobbo2001}
D.~Gobbo, M.~Napolitano, P.~Famouri, M.~Innocenti, Experimental application of extended kalman filter for sensor validation, Control Systems Technology, IEEE Transactions on 9 (2001) 376 -- 380.
\newblock \href {https://doi.org/10.1109/87.911389} {\path{doi:10.1109/87.911389}}.

\bibitem{Barrera}
J.~R. Barrera~Loredo, J.~Aguirre, L.~Ruano, Influence of high heels on walking motion: Gait analysis, Journal of applied biomechanics (12 2015).

\bibitem{Kim2013}
Y.~Kim, J.-M. Lim, B.~Yoon, Changes in ankle range of motion and muscle strength in habitual wearers of high-heeled shoes, Foot \& ankle international. / American Orthopaedic Foot and Ankle Society [and] Swiss Foot and Ankle Society 34 (2013) 414--9.
\newblock \href {https://doi.org/10.1177/1071100712468562} {\path{doi:10.1177/1071100712468562}}.

\bibitem{Maduabuchi}
J.~Maduabuchi, N.~Joseph, A.~Egwuonwu, A.~O. Ezeukwu, C.~Nwafulume, Effects of different heel heights on selected gait parameters of young undergraduate females, Page Header Logo Journal of Paramedical Sciences 3 (2012) 2008--4978.

\bibitem{Shang2020-nb}
J.~Shang, X.~Geng, C.~Wang, L.~Chen, C.~Zhang, J.~Huang, X.~Wang, A.~Yan, X.~Ma, Influences of high-heeled shoe parameters on gait cycle, center of pressure trajectory, and plantar pressure in young females during treadmill walking, J. Orthop. Surg. (Hong Kong) 28~(2) (2020) 2309499020921978.

\bibitem{patel2012wearable}
S.~Patel, H.~Park, P.~Bonato, L.~Chan, M.~Rodgers, A review of wearable sensors and systems with application in rehabilitation, Journal of NeuroEngineering and Rehabilitation 9~(1) (2012) 21.
\newblock \href {https://doi.org/10.1186/1743-0003-9-21} {\path{doi:10.1186/1743-0003-9-21}}.

\bibitem{muro2014gait}
A.~Muro-de-la Herran, B.~Garcia-Zapirain, A.~Mendez-Zorrilla, Gait analysis methods: An overview of wearable and non-wearable systems, highlighting clinical applications, Sensors 14~(2) (2014) 3362--3394.
\newblock \href {https://doi.org/10.3390/s140203362} {\path{doi:10.3390/s140203362}}.

\bibitem{caldas2017human}
R.~Caldas, M.~Mundt, W.~Potthast, F.~Buarque~de Lima~Neto, B.~Markert, Human motion analysis using wearable sensors: A review, IEEE Sensors Journal 17~(2) (2017) 394--409.
\newblock \href {https://doi.org/10.1109/JSEN.2016.2628340} {\path{doi:10.1109/JSEN.2016.2628340}}.

\bibitem{del2016free}
S.~Del~Din, A.~Godfrey, B.~Galna, S.~Lord, L.~Rochester, Free-living gait characteristics in ageing and parkinson's disease: Impact of environment and ambulatory bout length, Journal of NeuroEngineering and Rehabilitation 13~(1) (2016) 46.
\newblock \href {https://doi.org/10.1186/s12984-016-0154-5} {\path{doi:10.1186/s12984-016-0154-5}}.

\bibitem{sabatini2006quaternion}
A.~M. Sabatini, Quaternion-based strap-down integration method for applications of inertial sensing to gait analysis, Medical \& Biological Engineering \& Computing 44~(5) (2006) 432--441.
\newblock \href {https://doi.org/10.1007/s11517-006-0040-2} {\path{doi:10.1007/s11517-006-0040-2}}.

\bibitem{tao2012gait}
W.~Tao, T.~Liu, R.~Zheng, H.~Feng, Gait analysis using wearable sensors, Sensors 12~(2) (2012) 2255--2283.
\newblock \href {https://doi.org/10.3390/s120202255} {\path{doi:10.3390/s120202255}}.

\bibitem{[07]useofIMUs}
A.~G.-M. A.~I. Cuesta-Vargas, J.~M. Williams, The use of inertial sensors system for human motion analysis, Physical Therapy Reviews 15~(6) (2010) 462–473.

\bibitem{grewal2001kalman}
M.~S. Grewal, A.~P. Andrews, Kalman Filtering: Theory and Practice Using MATLAB, John Wiley \& Sons, 2001.

\bibitem{kalman1960new}
R.~E. Kalman, A new approach to linear filtering and prediction problems, Transactions of the ASME--Journal of Basic Engineering 82~(1) (1960) 35--45.
\newblock \href {https://doi.org/10.1115/1.3662552} {\path{doi:10.1115/1.3662552}}.

\bibitem{li2002survey}
X.~R. Li, V.~P. Jilkov, Survey of maneuvering target tracking. part v: Multiple-model methods, IEEE Transactions on Aerospace and Electronic Systems 39~(4) (2002) 1333--1364.
\newblock \href {https://doi.org/10.1109/TAES.2003.1251466} {\path{doi:10.1109/TAES.2003.1251466}}.

\bibitem{moe2004trunk}
R.~Moe-Nilssen, J.~L. Helbostad, Estimation of gait cycle characteristics by trunk accelerometry, Gait \& Posture 20~(3) (2004) 253--260.
\newblock \href {https://doi.org/10.1016/j.gaitpost.2003.10.002} {\path{doi:10.1016/j.gaitpost.2003.10.002}}.

\bibitem{kavanagh2008accelerometry}
J.~J. Kavanagh, H.~B. Menz, Accelerometry: A technique for quantifying movement patterns during walking, Gait \& Posture 28~(1) (2008) 1--15.
\newblock \href {https://doi.org/10.1016/j.gaitpost.2007.10.010} {\path{doi:10.1016/j.gaitpost.2007.10.010}}.

\bibitem{zijlstra2003spatio}
W.~Zijlstra, A.~L. Hof, Assessment of spatio-temporal gait parameters from trunk accelerations during human walking, Gait \& Posture 18~(2) (2003) 1--10.
\newblock \href {https://doi.org/10.1016/s0966-6362(02)00190-x} {\path{doi:10.1016/s0966-6362(02)00190-x}}.

\end{thebibliography}

\end{document}